\documentclass[twocolumn]{aastex701} 

\usepackage{graphicx} 
\usepackage{hyperref}
\usepackage{enumitem}
\usepackage{amssymb}
\usepackage{lipsum}
\usepackage{lineno}

\newcommand{\ltsima}{\stackrel{\textstyle <}{\sim}}
\newcommand{\simlt}{\scriptsize{\raisebox{-2pt}{$\ltsima$}}\normalsize}

\newcommand{\um}{\,$\mu$m}
\newcommand{\spherex}{SPHEREx}

\newcommand{\hto}{H$_2$O}

\newcommand{\asec}{$^{\prime\prime}$}

\newcommand{\ddeg}{$^{\rm o}$}
\newdimen\saa  \newdimen\sbb

\def\arcsec{\ifmmode {^{\scriptstyle\prime\prime}}
          \else $^{\scriptstyle\prime\prime}$\fi}
\def\arcmin{\ifmmode {^{\scriptstyle\prime}}
          \else $^{\scriptstyle\prime}$\fi}               
\def\parcs{\saa=.07em \sbb=.03em
     \ifmmode \hbox{\rlap{.}}^{\scriptstyle\prime\kern -\sbb\prime}\hbox{\kern -\saa}
     \else \rlap{.}$^{\scriptstyle\prime\kern -\sbb\prime}$\kern -\saa\fi}
\def\deg{\ifmmode^\circ\else$^\circ$\fi}

\def\o2{O$_2$}
\def\h2{H$_2$}
\def\co2{CO$_2$}
\def\ch3oh{CH$_3$OH}
\def\nh3{NH$_3$}
\def\av{$A_V$}
\def\cm3{cm$^{-3}$}

\begin{document}
\makebox[17cm][r]{\textcopyright 2025 all rights reserved.}

\linenumbers

\author[0000-0002-6025-0680]{Gary J. Melnick}
\email{gmelnick@cfa.harvard.edu}
\affiliation{Center for Astrophysics $|$  Harvard {\&} Smithsonian, 60 Garden St., Cambridge, MA 02138-1516, USA}

\author[0000-0002-5599-4650]{Joseph L. Hora}
\email{jhora@cfa.harvard.edu}
\affiliation{Center for Astrophysics $|$  Harvard {\&} Smithsonian, 60 Garden St., Cambridge, MA 02138-1516, USA}

\author[0000-0002-3993-0745]{Matthew L. N. Ashby}
\email{mashby@cfa.harvard.edu}
\affiliation{Center for Astrophysics $|$  Harvard {\&} Smithsonian, 60 Garden St., Cambridge, MA 02138-1516, USA}

\author[0000-0003-1841-2241]{Volker Tolls}
\email{vtolls@cfa.harvard.edu}
\affiliation{Center for Astrophysics $|$  Harvard {\&} Smithsonian, 60 Garden St., Cambridge, MA 02138-1516, USA}

\author[0000-0001-8064-2801]{Jaeyeong Kim}
\email{jaeyeong.kim@cfa.harvard.edu}
\affiliation{Korea Astronomy and Space Science Institute (KASI), 776 Daedeok-daero, Yuseong-gu, Daejeon 34055, Republic of Korea }
\affiliation{Center for Astrophysics $|$  Harvard {\&} Smithsonian, 60 Garden St., Cambridge, MA 02138-1516, USA}

\author[0000-0002-9548-1526]{Carey M. Lisse}
\affil{Johns Hopkins University, 3400 N Charles St, Baltimore, MD 21218, USA}
\affil{Johns Hopkins University, Applied Physics Laboratory, Laurel, MD 20723, USA}
\email{Carey.Lisse@jhuapl.edu}

\author[0000-0002-5158-243X]{Roberta Paladini}
\email{paladini@ipac.caltech.edu}
\affiliation{Infrared Processing Analysis Center, California Institute of Technology, Pasadena, CA 91125, USA}

\author[0000-0003-4990-189X]{Michael W. Werner}
\email{michael.w.werner@jpl.nasa.gov}
\affiliation{Jet Propulsion Laboratory, California Institute of Technology, 4800 Oak Grove Drive, Pasadena, CA, 91109, USA}

\author[0000-0003-3119-2087]{Jeong-Eun Lee}
\affil{Department of Physics and Astronomy, Seoul National University, 1 Gwanak-ro, Gwanak-gu, Seoul 08826, Korea}
\affil{SNU Astronomy Research Center, Seoul National University, 1 Gwanak-ro, Gwanak-gu, Seoul 08826, Republic of Korea}
\email{lee.jeongeun@snu.ac.kr}

\author[0009-0004-7886-9029]{Young-Jun Kim}
\affil{Department of Physics and Astronomy, Seoul National University, 1 Gwanak-ro, Gwanak-gu, Seoul 08826, Korea}
\email{yj\_kim1012@snu.ac.kr}

\author[0000-0002-5016-050X]{Miju Kang}
\affiliation{Korea Astronomy and Space Science Institute (KASI), 776 Daedeok-daero, Yuseong-gu, Daejeon 34055, Republic of Korea }
\email{mjkang@kasi.re.kr}

\author[0000-0002-5437-0504]{Yun-Ting Cheng}
\affiliation{California Institute of Technology, 1200 E. California Boulevard, Pasadena, CA 91109, USA}
\affiliation{Jet Propulsion Laboratory, California Institute of Technology, 4800 Oak Grove Drive, Pasadena, CA, 91109, USA}
\email{ycheng3@caltech.edu}

\author[0000-0002-5710-5212]{James~J.~Bock}
\affiliation{California Institute of Technology, 1200 E. California Boulevard, Pasadena, CA 91109, USA}
\affiliation{Jet Propulsion Laboratory, California Institute of Technology, 4800 Oak Grove Drive, Pasadena, CA, 91109, USA}
\email{jjb@astro.caltech.edu}

\author[0000-0002-4650-8518]{Brendan P. Crill}
\affiliation{Jet Propulsion Laboratory, California Institute of Technology, 4800 Oak Grove Drive, Pasadena, CA, 91109, USA}
\affiliation{California Institute of Technology, 1200 E. California Boulevard, Pasadena, CA 91109, USA}
\email{bcrill@jpl.nasa.gov}

\author[0000-0002-7471-719X]{Ari Cukierman}
\affiliation{California Institute of Technology, 1200 E. California Boulevard, Pasadena, CA 91109, USA}
\email{ajcukier@caltech.edu}

\author[0000-0001-7432-2932]{Olivier Dor\'{e}}
\affiliation{Jet Propulsion Laboratory, California Institute of Technology, 4800 Oak Grove Drive, Pasadena, CA, 91109, USA}
\affiliation{California Institute of Technology, 1200 E. California Boulevard, Pasadena, CA 91109, USA}
\email{olivier.dore@caltech.edu }

\author[0000-0002-9382-9832]{Andreas Faisst}
\affiliation{Infrared Processing Analysis Center, California Institute of Technology, Pasadena, CA 91125, USA}
\email{afaisst@caltech.edu}

\author[0000-0001-5812-1903]{Howard Hui}
\affiliation{California Institute of Technology, 1200 E. California Boulevard, Pasadena, CA 91109, USA}
\email{hhui@caltech.edu}

\author[0000-0002-2770-808X]{Woong-Seob Jeong}
\affiliation{Korea Astronomy and Space Science Institute (KASI), 776 Daedeok-daero, Yuseong-gu, Daejeon 34055, Republic of Korea }
\email{jeongws@kasi.re.kr}

\author[0000-0002-2523-3762]{Chul-Hwan Kim}
\affil{Department of Physics and Astronomy, Seoul National University, 1 Gwanak-ro, Gwanak-gu, Seoul 08826, Korea}
\email{chkim9407@snu.ac.kr}

\author[0000-0002-3808-7143]{Ho-Gyu Lee}
\affiliation{Korea Astronomy and Space Science Institute (KASI), 776 Daedeok-daero, Yuseong-gu, Daejeon 34055, Republic of Korea }
\email{hglee@kasi.re.kr}

\author[0000-0003-0894-7824]{Jae-Joon Lee}
\affiliation{Korea Astronomy and Space Science Institute (KASI), 776 Daedeok-daero, Yuseong-gu, Daejeon 34055, Republic of Korea }
\email{leejjoon@kasi.re.kr}

\author[0000-0001-5382-6138]{Daniel Masters}
\affiliation{Infrared Processing Analysis Center, California Institute of Technology, Pasadena, CA 91125, USA}
\email{dmasters@ipac.caltech.edu}

\author[0000-0002-5710-5212]{Chi~H.~Nguyen}
\affiliation{California Institute of Technology, 1200 E. California Boulevard, Pasadena, CA 91109, USA}
\email{chnguyen@caltech.edu}

\author[0009-0007-1206-9506]{Jinyoung Noh}
\affil{Department of Physics and Astronomy, Seoul National University, 1 Gwanak-ro, Gwanak-gu, Seoul 08826, Korea}
\email{kyla2001@snu.ac.kr}

\author[0000-0002-0070-3246]{Ji Yeon Seok}
\affiliation{Korea Astronomy and Space Science Institute (KASI), 776 Daedeok-daero, Yuseong-gu, Daejeon 34055, Republic of Korea }
\email{jyseok@kasi.re.kr}

\author[0000-0001-9842-639X]{Soung-Chul Yang}
\affiliation{Korea Astronomy and Space Science Institute (KASI), 776 Daedeok-daero, Yuseong-gu, Daejeon 34055, Republic of Korea }
\email{sczoo@kasi.re.kr}

\author[0000-0003-3078-2763]{Yujin Yang}
\affiliation{Korea Astronomy and Space Science Institute (KASI), 776 Daedeok-daero, Yuseong-gu, Daejeon 34055, Republic of Korea }
\email{yyang@kasi.re.kr}

\author[0000-0001-8253-1451]{Michael Zemcov}
\affiliation{School of Physics and Astronomy, Rochester Institute of Technology, 1 Lomb Memorial Dr., Rochester, NY 14623, USA}
\affiliation{Jet Propulsion Laboratory, California Institute of Technology, 4800 Oak Grove Drive, Pasadena, CA, 91109, USA}
\email{mbzsps@rit.edu}

\title{The SPHEREx Ices Investigation: An Overview}
\date{\today}


\begin{abstract}
SPHEREx is a NASA mission designed to perform an all-sky spectroscopic survey in the 0.75 -- 5\um\ wavelength range. Its primary science objectives are to investigate: (1) inflationary cosmology, (2) the history of galaxy formation, and (3) the abundance of molecular ices -- critical for prebiotic chemistry -- found on the surfaces of interstellar dust grains within planet-forming regions. This paper focuses on the third theme, the SPHEREx Ices investigation, for which SPHEREx is conducting a spectroscopic survey of nearly ten million preselected sources throughout the Milky Way and Magellanic Clouds to characterize their ice absorption features.  By selecting targets based on infrared color, spatial isolation, and brightness, the Ices Investigation secures high–signal-to-noise spectra across a broad range of astrophysical environments that are relatively free of spectral contamination. Rather than attempting to decompose each spectrum into its individual ice components, the Ices Investigation prioritizes accurate measurements of the integrated optical depths of key molecular ice absorption features. This approach enables statistically powerful correlation studies between ice abundances and environmental parameters -- including extinction, temperature, gas composition, radiation field strength, cosmic ray flux, and star formation activity.  The data pipeline developed for this purpose incorporates machine learning for continuum estimation, drawing on both SPHEREx and ancillary datasets. Ultimately, the expansive spectral archive produced by SPHEREx, combined with targeted follow-up from facilities like JWST, will transform our understanding of Galactic ice formation, evolution, abundance and their inheritance into planetary systems and prebiotic inventories.
\end{abstract}


\section{Introduction}

Interstellar and protoplanetary ices are key -- but still incompletely characterized -- reservoirs of volatile material that shape planetary compositions across sublimation snow lines and influence the potential for habitability within them. The remarkable chemical complexity of these ices arises from the formation and accretion of ices on dust grain mantles, governed by factors such as temperature, gas composition and density, far-ultraviolet (FUV) irradiation, cosmic-ray flux, and evolutionary timescales. Disentangling the relative contributions of these drivers to ice chemistry remains a central challenge. The overarching goal of the SPHEREx Ices Investigation is to advance our understanding of the mechanisms that determine interstellar ice compositions across environments -- from diffuse clouds to protoplanetary disks -- and to assess whether planet‑forming disks inherit their volatile inventories largely intact from their parent molecular clouds, or whether those volatiles are significantly reprocessed in situ.

Prior to the launch of the James Webb Space Telescope \citep[JWST;][]{Gardner2006}, our understanding of interstellar ices was largely built on data from ground-based observatories, such as United Kingdom Infrared Telescope \citep[UKIRT;][]{Chiar1995}, NASA’s Infrared Telescope Facility \citep[IRTF;][]{Chiar2011} and ESO's Very Large Telescope \citep[VLT;][]{Pontoppidan2003,Pontoppidan2006}, along with space-based missions including the Infrared Space Observatory \citep[ISO; e.g.,][]{Gibb2004}, the Spitzer Space Telescope \citep{Boogert2008}, and Japan’s AKARI mission \citep{Murakami2007}. Collectively, these facilities obtained approximately 400 near- and mid-infrared ice spectra. While these observations securely established interstellar ices as major reservoirs of molecules relevant to prebiotic chemistry and planetary habitability, the limited sample size precluded disentangling the roles of different physical and chemical processes shaping ice composition.

The advent of JWST has opened new and transformative opportunities for the study of interstellar and protoplanetary ices. With its unmatched sensitivity, broad wavelength coverage (1–28\um), and wide range of spectral resolving powers (R$\:\equiv \lambda/\Delta\lambda =$ 30–3700), JWST is producing the most detailed ice spectra to date and has already amassed several hundred spectra. However, whether this dataset is sufficient to identify the dominant mechanisms that govern ice composition and distribution across varied environments remains an open question. Given the diversity of conditions in which ices form and evolve, a far larger statistical sample will almost certainly be needed to fully unravel the processes that determine their composition and abundance.

The SPHEREx mission \citep{Bock2025} will provide a valuable, complementary approach to advancing our understanding of cosmic ices. In particular, the Ices Investigation is dedicated to exploring the relationship between ices in interstellar clouds and those in planet-forming disks by measuring the composition and relative abundances of key ices -- e.g., water (\hto), methanol (CH$_3$OH), carbon dioxide (\co2), XCN (most commonly attributed to OCN$^-$ ions), carbon monoxide (CO), and 
carbonyl sulfide (OCS) -- across an unbiased and diverse sample of at least 20,000 targets in the Milky Way and Magellanic Clouds.  The minimum of 20,000 targets reflects the science requirements established during SPHEREx mission development. However, this number represents only a lower bound; as discussed below, we expect the actual yield to be significantly higher.

Through its all-sky survey, SPHEREx is expected to obtain absorption spectra along nearly 10 million pre-selected lines of sight \citep[see][]{Ashby2023}. These observations will probe environments ranging from diffuse clouds to dense protoplanetary disks, with a total extinction in the V band, \av, measured in magnitudes, ranging from approximately \av$\sim\,$2 to more than 100. By leveraging statistically robust samples within each environment, we will be able to examine how both the integrated optical depth of each ice species and the variations among these ice features correlate with the prevailing physical conditions. Equally important, a lack of clear correlations would provide valuable insight, pointing to additional, but as yet unidentified, processes that regulate the composition and evolution of astrophysical ices.

Together, JWST and SPHEREx can provide both the depth and breadth of ice spectroscopy needed to move beyond detection and toward a more comprehensive picture in which the composition and distribution of ices can be understood in terms of the underlying environment that shape them.

This paper presents an overview of the SPHEREx Ices Investigation and is organized into six sections. Section 2 explores the role of ices in the interstellar medium (ISM) and protoplanetary disks (PPDs). Section 3 provides a summary of the SPHEREx instrument and describes how spectra are acquired. Section 4 details the criteria for selecting suitable ice targets for observation. Section 5 discusses the primary challenges inherent in our approach and outlines the strategies developed to overcome them. Section 6 introduces the statistical methods employed to analyze the SPHEREx dataset, explaining our focus on each feature's integrated ice optical depth rather than possible ice mixture fits. Finally, Section 7 describes the Ices data pipeline and the resulting data products.

Finally, throughout this paper, for simplicity and clarity, we refer to various ice absorption features only by their dominant constituent -- e.g., \hto, \co2, and CO -- even though, in most cases, each feature arises from a mixture of ices. The complexities of these mixtures are discussed further in Section~6.

\section{The Crucial Role of Ices}

Studies of our Solar System provide compelling evidence that ices are abundant beyond roughly 3 AU. A large fraction \citep[e.g.,][]{Podolak2020} of the “ice giants” Uranus and Neptune are composed of water, methane, and ammonia ice. Among Jupiter’s satellites, Ganymede and Callisto are especially ice-rich -- surpassing even Europa in their ice content \citep[e.g.,][]{Schubert2004}. Europa, which is slightly smaller than Earth’s Moon, contains twice as much water as all of Earth’s oceans, a consequence of its formation in a region of the solar system where water ice is stable.  Saturn’s system is similarly dominated by ice: Titan, Enceladus, and Triton all harbor substantial reservoirs, while Saturn’s rings consist of more than 90\% water ice \citep{Hussmann2006,Tobie2006,Schubert2007,Cuzzi2010}. At greater distances, the Kuiper Belt and Oort Cloud host millions of icy bodies that serve as the primary sources of comets, with total masses of approximately 0.01–0.1 Earth masses \citep{Gladman2008} and 1–10 Earth masses \citep{Dones2004}, respectively.

The Solar System is unlikely to be unique in this regard. Although direct detections of icy bodies around other stars remain scarce, ice is thought to play a central role during the early stages of planet formation and in setting the composition of planets beyond the major snow lines \citep[e.g., \hto, \co2, CO;][]{Oberg2011,Oberg2023}. The connection between ices in protoplanetary disks and those inherited from parent molecular clouds is an area of active investigation. Nevertheless, it is already evident that ice formation begins early in the evolution of molecular clouds \citep[e.g.,][]{McClure2023}.

\begin{figure}
    \centering
    \includegraphics[width=1.0\linewidth]{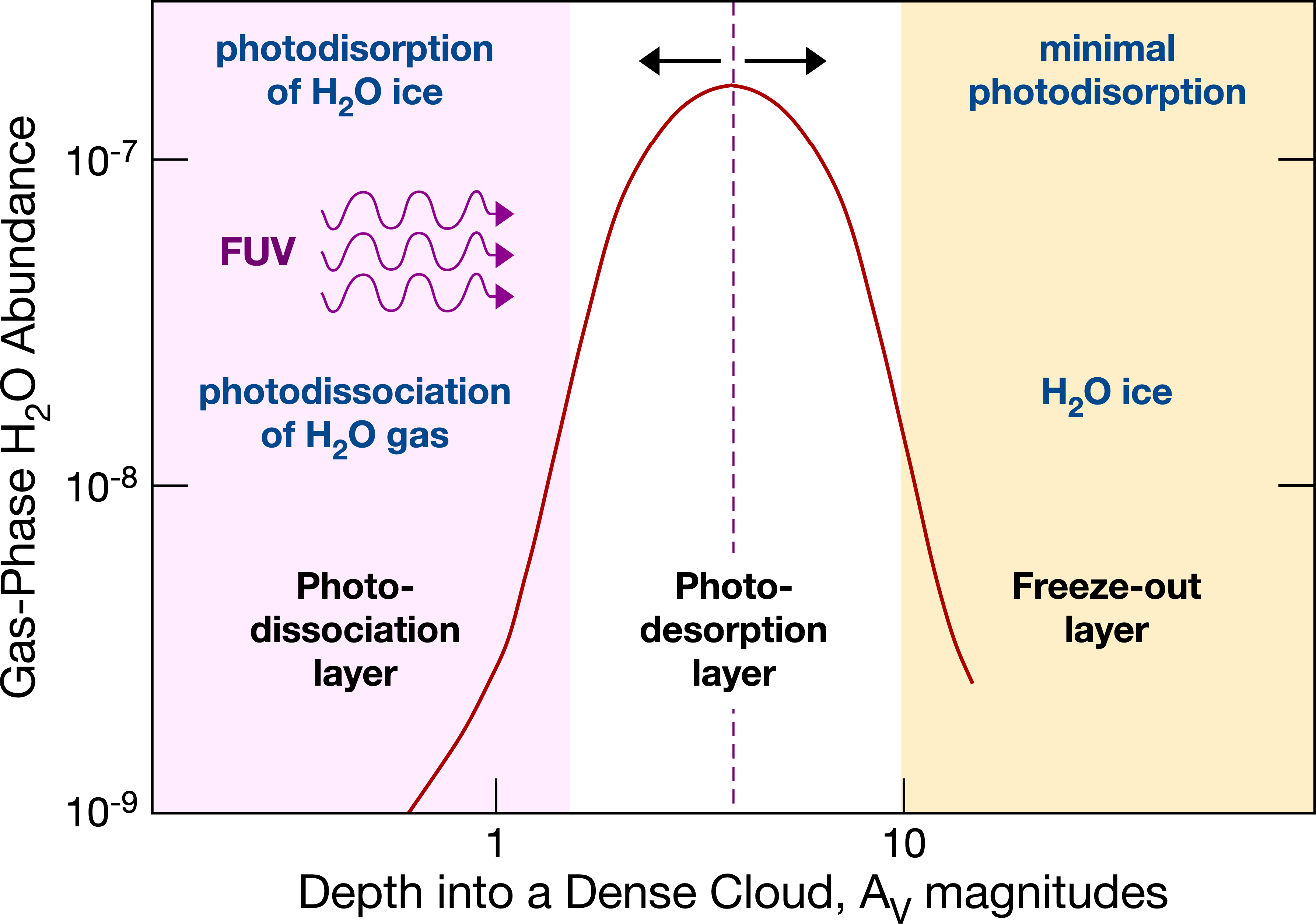}
    \caption{Schematic illustration of the transition from gas-phase water to water ice. At the surface of a molecular cloud, far-ultraviolet (FUV) radiation efficiently photodissociates water molecules -- regardless of whether they form in the gas or are produced on dust grains and subsequently photodesorbed.  Deeper into the cloud, the FUV field is attenuated but still present; in this partially shielded zone, photodesorbed water molecules survive longer, leading to an increased abundance of gas-phase water. Toward the cloud interior, the FUV field is almost fully suppressed, photodesorption becomes inefficient, and with grains remaining cold ($\simlt$\,25 K), thermal desorption is negligible. Under these shielded, low-temperature conditions, water predominantly exists as ice on grain surfaces, and the abundance of water ice greatly exceeds that of gas-phase water.}
    \label{fig:notional}
\end{figure}

The Submillimeter Wave Astronomy Satellite (SWAS) was the first observatory to access the 556.936 GHz 1$_{10}$ - 1$_{01}$ ground-state transition of ortho-\hto\  \citep{Melnick2000}. With an upper-state energy, $E_u/k\,=\,$27\,K, this line was well suited for probing gas-phase water in cold molecular clouds, with characteristic temperatures of 10\,-\,20\,K. Spectra acquired for several thousand lines of sight toward nearly 400 Galactic molecular clouds revealed that, aside from hot gas associated with molecular outflows, the measured gas-phase \hto\ column density did not scale with the inferred \h2\ column density. Instead, gas-phase \hto\ column densities remained confined to a narrow range, independent of the total \h2\ content. This led to the conclusion that water vapor resides primarily in the FUV-illuminated regions of cloud surfaces \citep{Melnick2011}, while gas-phase water abundances drop sharply in the well-shielded interiors -- a result later confirmed by Herschel observations \citep{Melnick2020, Dishoeck2021}.  This is shown schematically in Figure~\ref{fig:notional}.

\begin{figure*}
    \centering
    \includegraphics[width=0.90\linewidth]{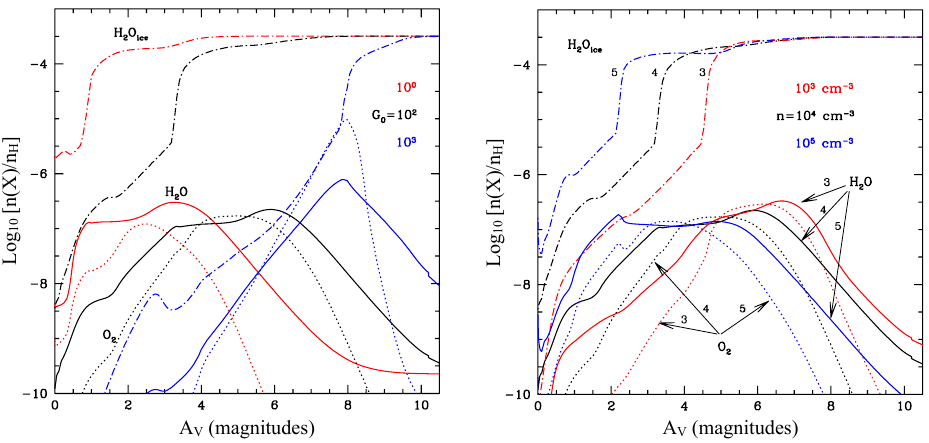}
    \caption{({\em Left}) Predicted abundances of gas-phase \hto\ (solid), \hto\ ice (dot–dashed), and \o2\ (dotted) as a function of $A_V$ for a cloud with $n_\mathrm{H} = 10^4$ \cm3, exposed to FUV field strengths, $G_0$, of 1, 100, and 1000 times the Solar neighborhood value. Increasing $G_0$ shifts the abundance profiles to larger $A_V$. While the freeze-out depth depends on $G_0$, the total \hto\ ice column is unchanged. The enhanced \o2\ peak at $G_0 = 1000$ results from thermal desorption of atomic O from warm grains, which suppresses \hto\ ice formation and leaves more O in the gas phase. ({\em Right}) Dependence on gas density for $n = 10^3$, $10^4$, and $10^5$ \cm3\ at $G_0 = 100$. Both the threshold $A_V$ for ice formation and the $A_V$ of the \hto\ and \o2\ peaks increase with $G_0/n$, though the peak abundances are largely independent of density. Arrows indicate $\log n_\mathrm{H}$. \citep[After][]{Hollenbach2009}.}
    \label{fig:Hollenbach}
\end{figure*}

SWAS also carried a receiver for the 487.249 GHz 3,1 - 3,2 transition of \o2\ ($E_u/k\,=\,$26\,K), observed simultaneously with \hto\ but never detected with confidence. The unexpectedly low abundance of \o2, together with the suppressed abundance of the gas-phase \hto\ in cloud interiors, is consistent with models in which most of the elemental oxygen not in CO is locked in \hto\ ice.  \citet{Hollenbach2009} predict that at a depth into the cloud of 3–8 magnitudes of visual extinction (depending on gas density and the strength of the external FUV field; see Figure~\ref{fig:Hollenbach}), more than 99\% of the water resides in ice -- a prediction SPHEREx was designed to test.

CO is abundant in the gas phase of molecular clouds but freezes onto dust grains when temperatures fall below $\sim\,$25\,K. Observations toward pre-stellar cores and cold cloud interiors show a sudden decline in gaseous CO emission due to freeze-out \citep[e.g.,][]{Bacmann2002,Carolan2008,Christie2012,Smith2025}. Freeze-out, and the resulting loss of gas-phase CO as a major coolant, can affect both the thermal balance and fragmentation of molecular clouds \citep[e.g.,][]{Hocuk2014}.

In contrast, \co2\ ice is not thought to form via direct freeze-out of gas-phase \co2, but rather through grain-surface reactions \citep[e.g.,][]{Gibb2004,Oba2010}.  The dominant pathway involves reactions between CO and O atoms on cold dust grains, proceeding via either Langmuir–Hinshelwood or Eley–Rideal heterogeneous catalysis reactions depending on temperature \citep[][and references therein]{Roser2001, Minissale2013, Herbst2021}. Alternatively, CO may react with OH radicals to produce \co2\ \citep[e.g.,][]{Ioppolo2011, Molpeceres2023}. At higher temperatures (up to $\sim\,$150\,K), \co2\ can also form within water-rich carbon-bearing ices subjected to UV irradiation or energetic particle bombardment \citep{Potapov2023}.

Methanol (CH$_3$OH), methane (CH$_4$), ammonia (NH$_3$), and other complex organic ices have also been detected in the cold, dense regions of the ISM \citep[e.g.,][]{McClure2023}. These molecules are primarily formed through surface reactions on grains, enabled by low temperatures and the accretion of simple species such as CO, \hto, and N. Methanol, one of the most abundant complex organics detected in interstellar ices, is a particularly important building block, as its presence, along with H$_2$CO (formaldehyde), signals active grain-surface chemistry and serves as a precursor to larger prebiotic molecules. Methane and ammonia ices form alongside methanol and water \citep{Boogert1997,Fedoseev2015}, reflecting different chemical pathways and environmental histories -- such as the influence of atomic hydrogen and available carbon or nitrogen sources. Detections of these ices reveal that complex organic species can become relatively abundant, especially in regions shielded from FUV radiation. Their subsequent processing by FUV photons and cosmic rays may drive the formation of even larger and more complex organic molecules, like PAHs, tholins, and amino acids, establishing a chemical link between interstellar ices and the raw materials of planetary systems, comets, and potentially, the origins of life \citep[e.g.,][]{Cruikshank2019}.

\setcounter{table}{0}

\section{The SPHEREx Instrument and Assembly of Spectra}

\begin{figure}
    \centering
    \vspace{2mm}
    \includegraphics[width=1.0\linewidth]{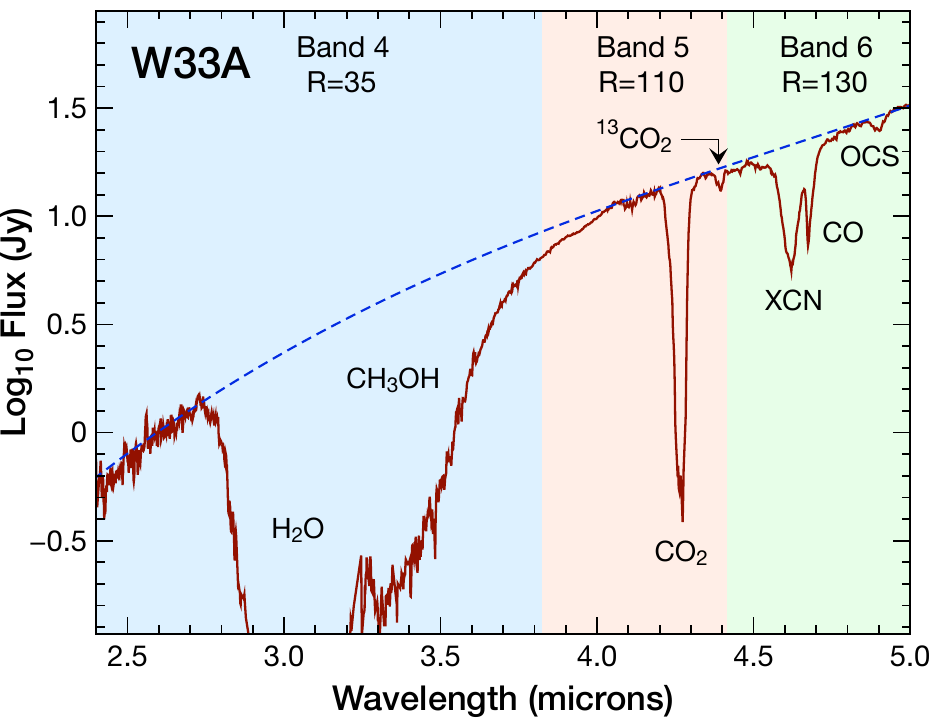}
    \caption{ISO Short Wavelength Spectrometer (SWS) spectrum of the Galactic source, W33A, an extremely embedded high-mass young stellar object \citep{Gibb2004}. This source exhibits strong ice absorption features due to \hto, CH$_3$OH, \co2, XCN, CO, and weaker $^{13}$CO$_2$ and OCS features.  Also shown are the SPHEREx Bands 4, 5, and 6, along with their respective spectral resolving powers, $\lambda/\Delta\lambda$ (see Table~1).}
    \label{fig:W33}
\end{figure}

SPHEREx is equipped with a 20-cm telescope and a wide-field (3.5\ddeg\,$\times$\,11.5\ddeg) imaging spectrometer that covers the near-infrared wavelength range from 0.75 to 5\um\ \citep{Korngut2025}. Spectra are produced using linear variable filters (LVFs), whose transmitted wavelength varies linearly along one physical dimension of the filter. By scanning the telescope and imaging different regions of each LVF onto the detectors, we obtain continuous spectral coverage across six bands, with resolving powers ($R \equiv \lambda/\Delta\lambda$) ranging from 35 to 130 (see Table~1). These spectral resolutions were selected to be well matched to the goals of the Ices Investigation, which aims to detect and spectrally separate broad solid-state ice absorption features -- such as \hto, CH$_3$OH, \co2, XCN, CO, and OCS, as shown in Figure~\ref{fig:W33} -- as well as signatures from both aromatic and aliphatic polycyclic aromatic hydrocarbons (PAHs) and starlight extinction.

The SPHEREx instrument utilizes six 2K\,$\times$\,2K Hawaii-2RG detector arrays, divided into three short-wavelength and three long-wavelength bands by a dichroic beam splitter. SPHEREx obtains spectra from multiple pointed exposures separated by both small and large slews. Successive exposures trace an approximate great circle 90\ddeg\ from the Sun. Two detector arrays observe the same sky region simultaneously (Bands 1 and 4, 2 and 5, 3 and 6), each sampling different wavelengths. As the telescope steps and integrates, sources moves across all six arrays, yielding a complete 0.75–5.0\um\ spectrum for every 6\farcs15 × 6\farcs15 pixel in the field. Because of this unique observing strategy, spectra are reconstructed by ground‑system software rather than onboard the instrument.

Since a full spectrum for any given source is built up gradually over the course of days or weeks (as the telescope scans the sky), accounting for possible source variability is important. This issue is addressed in Section~5.2.

Over its prime two-year mission, SPHEREx will conduct four complete all-sky surveys. With each successive survey, additional spectral samples will be added, yielding at least 408 independent spectrophotometric measurements per target. For further details, including further discussion of the instrument and survey strategy, see \citet{Bock2025} and \citet{Bryan2025}.

\begin{table}
\begin{center}
\rule{0mm}{5mm}
~~~~~~~~~TABLE 1.~SPHEREx Bands and \\
~~~~~~~~~Spectral Resolving Powers \\*[-3.5mm]
\begin{tabular}{ccc} \\ \hline
\rule{0mm}{4mm} &  & Spectral \\*[-1mm]
  & ~~~Wavelength~~~  &  Resolving  \\*[-1mm]
~Band~ & Range & Power, $R$ \\*[-1mm]
  &  (\um)  &  ($\lambda/\Delta\lambda$) \\*[1mm] \hline
\rule{0mm}{4mm}1 & 0.75 -- 1.11  & ~~41 \\
2  & 1.11 -- 1.64  & ~~41 \\
3  & 1.64 -- 2.42  & ~~41 \\
4  & 2.42 -- 3.82  & ~~35 \\
5  & 3.82 -- 4.42  & 110 \\
6  & 4.42 -- 5.00  & 130 \\*[1mm] \hline
\end{tabular}
\end{center}
\label{bandres}
\end{table}

\section{Ices Investigation Targets}

The \spherex\ Level 1-3 (L1-L3) data pipeline \citep{Akeson2025} automatically creates spectra for celestial targets, but only at predefined locations.  For this reason, the \spherex\ science team identified suitable targets of all kinds relevant to the three main mission science themes \citep[see][]{Bock2025}, and compiled their coordinates in the \spherex\ Reference Catalog \citep{Yang2026}.  As shown in Figure~\ref{fig:W33}, near-infrared ice bands are detected through their line-of-sight absorption of background starlight. The subset of the Reference Catalog comprising suitable background sources for our absorption studies is called the \spherex\ target List of ICE Sources -- hereafter referred to as SPLICES targets \citep[see][]{Ashby2023}.

To identify suitable background sources, we drew upon two
well-characterized all-sky survey photometric databases, ALLWISE \citep{Cutri2014} and the Two Micron All Sky Survey \citep[2MASS;][]{Skrutskie2006}.  The Wide-field Infrared Survey Explorer \citep[WISE;][]{Wright2010} surveyed the entire sky at 3.4, 4.6, 12, and 22\,\um, achieving angular resolutions of 6\farcs1 and 6\farcs4 at 3.4 and 4.6\,\um, respectively, with an astrometric precision better than 0\farcs15 for high signal-to-noise sources. 2MASS likewise covered the entire sky in the $J$, $H$, and $K_s$ bands.  Together, 2MASS and ALLWISE furnish the cataloged sources from which SPHEREx ice absorption targets were selected. 

\begin{figure*}
    \centering
    \includegraphics[width=0.99\linewidth]{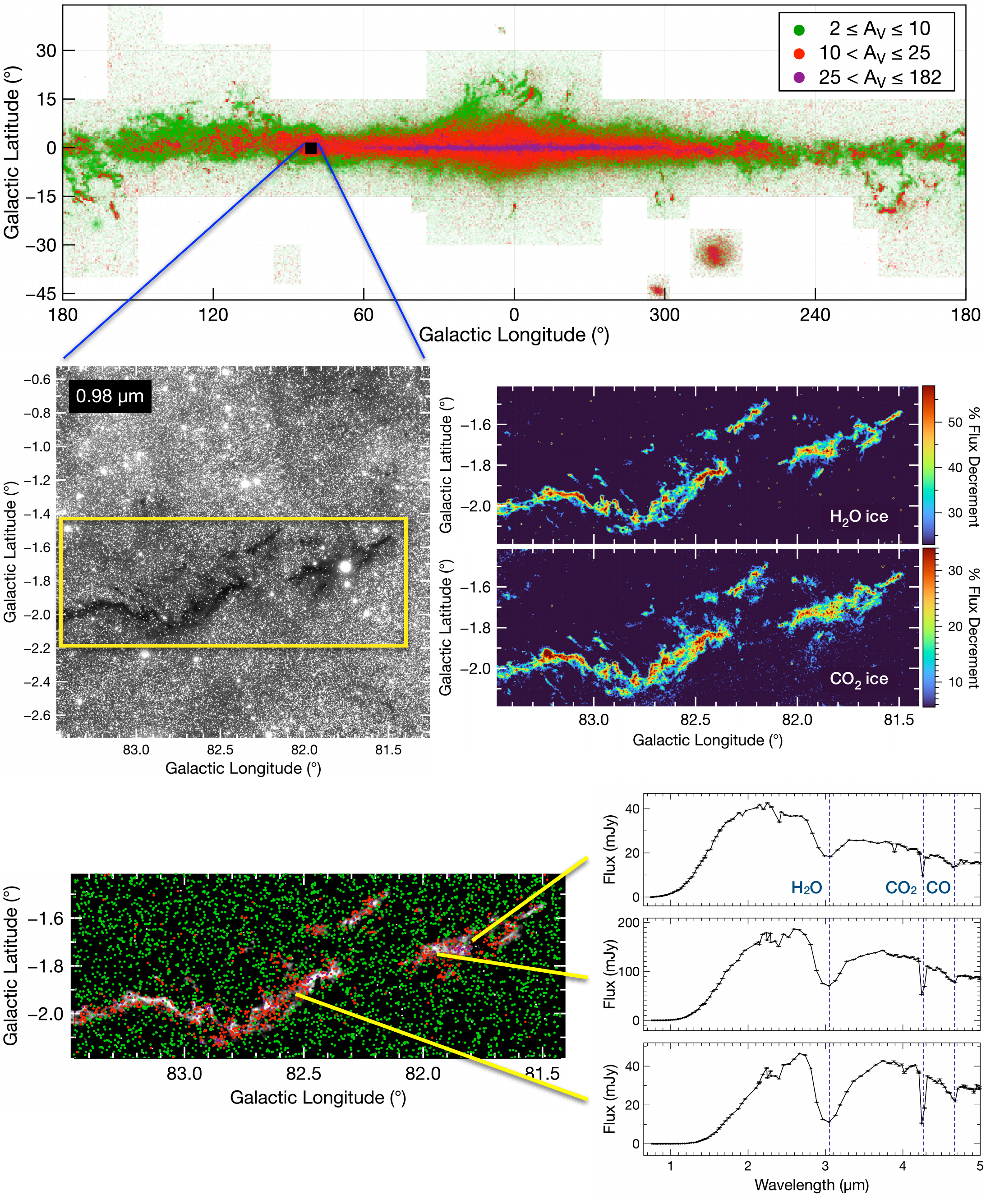}
    \caption{({\em Top}) Distribution of the more than 9.9 million pre‑selected SPHEREx ice targets in the Milky Way and Magellanic Clouds, color‑coded by estimated foreground dust extinction (\av in magnitudes; \citet{Ashby2023}). ({\em Middle left}) SPHEREx greyscale mosaic of the Cygnus region at 0.98\um; the yellow rectangle indicates the area expanded to the right for \hto‑ice and \co2‑ice. ({\em Middle right}) Percent flux decrement, 100 $\times$ (continuum-flux)/continuum, at 3.0\um\ (\hto‑ice) and 4.28\um\ (\co2‑ice). ({\em Lower left}) Greyscale mosaic of the \hto‑ice with SPLICES targets overlaid, using the same foreground‑extinction color scale as in the top panel. ({\em Lower right}) SPHEREx spectra for three of the 4,280 SPLICES targets within this region.  The small dip in the spectra at 2.4\um\ is an artifact caused by the instrument dichroic and will be automatically removed in future data processing.}
    \label{fig:Targets}
\end{figure*}

\begin{figure}
    \centering
    \includegraphics[width=0.98\linewidth]{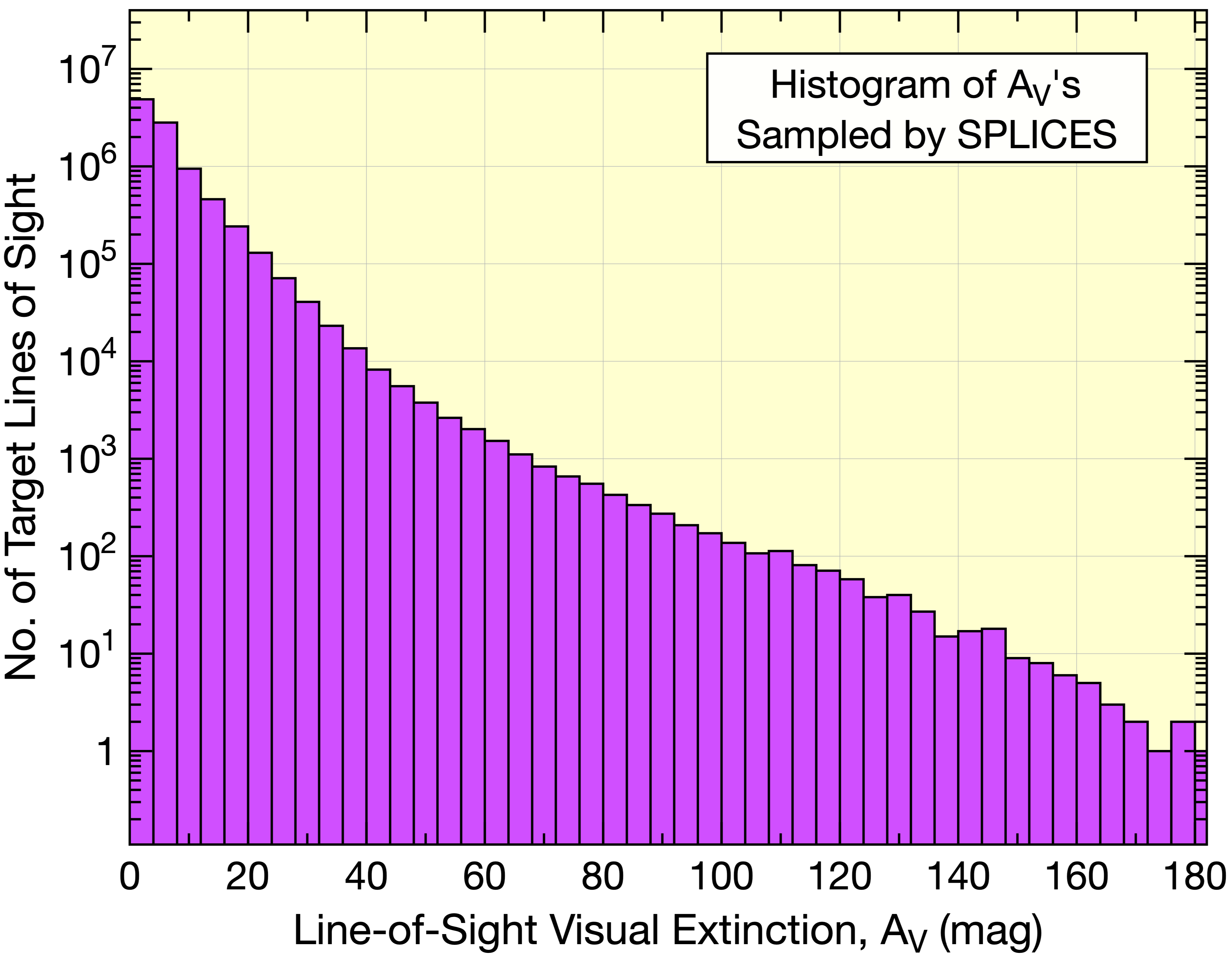}
    \caption{The distribution of line-of-sight dust visual extinctions toward SPLICES targets in \av\ terms, extrapolated roughly from the following relation: $A_K = 0.918\,(H - [4.5] - 0.08)$ \citep{Majewski2011}.  Here the WISE $W2$ magnitude has been substituted for the IRAC 4.5\um\ magnitude because it is available for all SPLICES targets.  $A_K/Av = 0.112$ \citep{Rieke1985} was assumed in order to derive the \av\ estimates.  In cases where the objects were detected in the 2MASS $Ks$ band but not in $H$, a similar relation was used to determine \av\ from the $Ks-$[4.5] color.  More than 83,000 SPLICES targets have an estimated line-of-sight \av\ extinction greater than 30 mag. These \av\ values are highly uncertain and should be regarded only as rough indicators of the total foreground dust extinction -- i.e., that SPLICES targets are either intrinsically red or likely to exhibit molecular absorption features in the SPHEREx bands. Detailed modeling and spectral typing will yield significantly more accurate \av\ estimates (see Section~5.1).
}
    \label{fig:avs}
\end{figure}

Targets for the Ices Investigation were selected using a variety of strategies designed to maximize both the number and quality of ice absorption detections. This approach is expected to yield numerous high signal-to-noise ratio (SNR) measurements of H$_2$O, CO$_2$, CO, and other ice features, providing substantial statistical power due to the large sample size. In addition, the target list was intentionally constructed to encompass a broad range of physical conditions, both intrinsic to the sources and within their surrounding environments, enabling comprehensive studies of how these factors influence ice composition and abundance.

The majority of SPLICES targets were chosen according to three selection criteria.  First, they were required to have red colors as an indicator of probable significant foreground extinction.  Specifically, infrared color (measured in the 2MASS H and WISE W2 bands) was used to estimate the line-of-sight extinction, following  \citet{Majewski2011}. We used a threshold of H-W2$\,>\,$0.324 mag, which corresponds roughly to $A_V>2$ for main-sequence stars, i.e., the level of shielding from UV radiation thought necessary for ices to freeze out onto grains (Hollenbach et al.\ 2009).  Two additional color cuts were also used to identify more deeply embedded sources, for which $H$- or $K_s$-band emission respectively might be faint or undetected, $K_s - W1 > 0.55$ and $W1-W2>0.1$.  By construction, then, all the SPLICES targets are detected in $W2$, and the vast majority are likewise detected by 2MASS, i.e.~98\%.

Second, SPLICES sources were required to be spatially isolated.  This criterion was introduced to avoid spectroscopic confusion arising from multiple targets within a single SPHEREx beam.  Accordingly, SPLICES targets were required to be free of neighboring sources within 6\farcs2 that contributed $\ge 1$\% of the target's flux at $H$, $K_s$, $W1$, and $W2$, where the ice features of interest lie.  Third and finally, SPLICES sources were required to be relatively bright.  To ensure reliable detections of spectral features, only sources brighter than $W2=11.96$\,Vega mag were included; at this level SPHEREx is expected to yield SNR$\ge50$ per spectral channel in Bands 5 and 6.

The SPLICES sample includes stars from both the Milky Way and the Magellanic Clouds. In total, the Ices Investigation has assembled nearly $10^7$ individual targets across these galaxies. The spatial distribution of these targets and a sample of SPHEREx spectra associated with each target is illustrated in Figure~\ref{fig:Targets}.  These targets also sample a wide range of \av's, as shown in Figure~\ref{fig:avs}. Note that the SPLICES targets are used to investigate ice properties along individual sightlines, rather than to construct spatially contiguous maps of the ice distribution. For the latter, we instead use the spatially continuous diffuse infrared background, as illustrated in the middle-right panel of Fig.~4. The SPLICES targets, by probing the ice content toward background sources, provide a more reliable measure of this content than the diffuse infrared background, for which the detectable ice content is limited to optical depths of $\tau_{\rm ice} \sim\,$1. Nevertheless, the diffuse infrared background is sufficient for determining the spatial distribution of the ice, and this procedure is described in detail in \citet{Hora2026a}.

A variety of additional sources were incorporated into the SPLICES target list to address a range of programmatic objectives. These supplementary targets include several hundred bright circumstellar disks, isolated stars from the Pleiades, Hyades, and M67 open clusters, Galactic halo stars, and stars near the North Ecliptic Pole 
used for flux-calibrating Spitzer's Infrared Array Camera.  Many of these objects were selected not for their potential to show ice features, but rather for qualities already documented in the literature such as minimal foreground dust or existing stellar spectral models \citep[e.g., the LAMOST and ATLAS stars;][]{Cui2012, Ji2023} 
which could be referenced by our team in the future to validate our own spectral modeling efforts.

A more detailed description of the SPHEREx target list is presented in \citet{Ashby2023}, including descriptions of the various types of ancillary photometry, astrometry, and variability data (none of which was used for target selection, but which is likely to help with interpretation of the SPHEREx spectra).  For sources that have it, for example, the ancillary data includes Spitzer/IRAC photometry, Gaia photometry and astrometry, and Pan-STARRS photometry. For further details, readers are encouraged to consult the SPLICES release notes maintained at the NASA/IPAC Infrared Science Archive (IRSA) 
\citep{https://doi.org/10.26131/irsa554},
which provide comprehensive descriptions of recent additions to the target list, and which will be updated to reflect any future modifications as well.

\section{Challenges and Solutions}

Below, we summarize the primary challenges involved in extracting and interpreting reliable spectra from the full-sky SPHEREx survey. Further details can be found in the Appendix.

\subsection{Stellar Typing}

A central aim of the Ices Investigation is to measure ice absorption features toward a large and diverse set of background stars at various stages of stellar evolution and to correlate these absorption measurements with the intervening dust extinction. Distinguishing between spectral features that originate in the background stars themselves and those due to foreground ices is therefore essential. This is achieved using advanced spectral modeling to remove stellar-intrinsic features from the data. As a first step, a machine-learning approach has been employed to classify millions of SPHEREx targets -- using broad photometric bands, sources are assigned to classes such as young stellar objects (YSOs), asymptotic giant branch (AGB) stars, active galactic nuclei (AGNs), or main-sequence (MS) stars \citep{Lakshmipathaiah2023}. These classifications, now publicly available, help prioritize targets for reliable ice measurements. Looking ahead, even greater accuracy is expected as new methods \citep[such as {\tt uberMS};][]{Pass2025} leverage the full SPHEREx spectral range to simultaneously determine intrinsic and extrinsic stellar parameters.

\subsection{Source Variability}

Source variability poses another challenge for robust measurement of ice features, because SPHEREx spectra are assembled from exposures spanning several days to weeks. Significant variability on these timescales can introduce artificial features or mask true absorption signals. 
To address this, multiple methods are used to flag variable sources. We have used the NEOWISE database to search for variability in the SPLICES targets, as described in \citet{Ashby2023}. We also performed a cross-match with the catalog of Gaia-identified variables \citep{Eyer2023} for the subset of targets that have Gaia detections. The preprocessor also performs multiple tests to detect variability in the SPHEREx observations themselves, including checking for discontinuities in the spectra and flux variations above what is expected from the measurement uncertainties in observations taken at different times at the same wavelength. Synthetic photometry is calculated from the SPHEREx spectra in the 2MASS and WISE bands and compared to archival data to identify long-term variability.
Approximately 22\% of SPLICES targets are flagged as variable and will either be excluded from the Ices analysis or subjected to additional processing -- depending on their variability characteristics -- to enable accurate determination of the IODs of their ice features.

\subsection{Number of Clouds Along the Line of Sight}

Accurately relating observed ice features to specific cloud environments requires knowing how many clouds lie along a given sightline. When multiple clouds are present, each can contribute absorption to the emergent spectrum, making interpretation more complex. To estimate the number of clouds along the lines of sight to ice targets, we used a method based on velocity-resolved CO spectroscopy, which can distinguish individual clouds by their unique Doppler shifts. 

Based on our analysis, approximately 94\% of the SPLICES targets are covered by the \citet{Dame2001} CO J=1–0 survey. Of these, 36\% have sightlines passing through a single molecular cloud, 24\% do not intersect any CO-detected cloud due to low extinction (i.e., \av$\, <\,$3–4 mag), and the remaining 40\% intersect two or more clouds along the line of sight. Single-cloud sightlines are relatively uniformly distributed across Galactic regions with detected CO emission, with one notable exception: in the mid-plane region ($-$50\ddeg$\,< \ell <\:$50\ddeg~and $\mid b\mid\,<\,$1\ddeg), all sightlines intersect multiple clouds. For details see Appendix A.3.1.

We cross-validated this approach for a subset of sightlines using 3D dust extinction maps, where clouds appear as sharp increases in extinction with distance. For our initial analysis, we will prioritize sightlines intersecting a single cloud, since these are simpler to interpret. Further details are provided in Appendix A.3.2.

\subsection{Radiative Transfer Modeling for Protoplanetary Disks}

Protoplanetary disk spectra present a unique challenge because of the disks’ complex three-dimensional structure and the spatially unresolved nature of SPHEREx measurements. To interpret these spectra, radiative transfer models are used to simulate how light interacts with dust, gas, and ices in the disks, allowing key physical and chemical parameters to be inferred.

For each ice absorption feature, integrated optical depths (see next section) will be cataloged so that researchers can apply their preferred radiative transfer models. The SPHEREx project will also exploit recent advances in disk modeling, incorporating zone‑dependent ice opacities and freeze‑out using tools such as RADMC‑3D and OpTool. A comprehensive grid of model spectra will be precomputed, enabling efficient fitting of thousands of observed spectra and robust extraction of disk properties and ice abundances.  Appendix section A.4 provides further details.

This modeling framework will not only constrain the structure and composition of disks, but also enable detailed, physically motivated studies of ice evolution across diverse disk environments. The resulting model parameters and derived properties will likewise be cataloged.

\section{Integrated Optical Depths and A Statistical Approach}

Interstellar and protoplanetary ices are almost invariably composed of molecular mixtures, dominated by \hto, \co2, and CO, along with smaller amounts of species such as \ch3oh, CH$_4$, \nh3, XCN, OCS, and others \citep[e.g.,][]{McClure2023, Rocha2024, Rocha2025}. The relative abundances of these constituents can vary widely from one line of sight to another, making it computationally challenging to determine a unique ice composition for each of nearly 10 million SPLICES targets. Accurately estimating the column densities of individual ice species is further complicated by two additional factors. First, model fits to observed ice absorption features, constructed by superposing laboratory ice-mixture spectra, are often non-unique; for instance, several different ice mixtures can reproduce the JWST-observed 4.38\um\ $^{13}$CO$_2$ absorption feature equally well \citep{Brunken2024a}. Second, absorption features beyond 5\um\ -- outside the SPHEREx wavelength range -- are often critical for discriminating among possible ice mixtures that may account for the features observed at shorter wavelengths.

\begin{figure}
    \centering
    \includegraphics[width=1.0\linewidth]{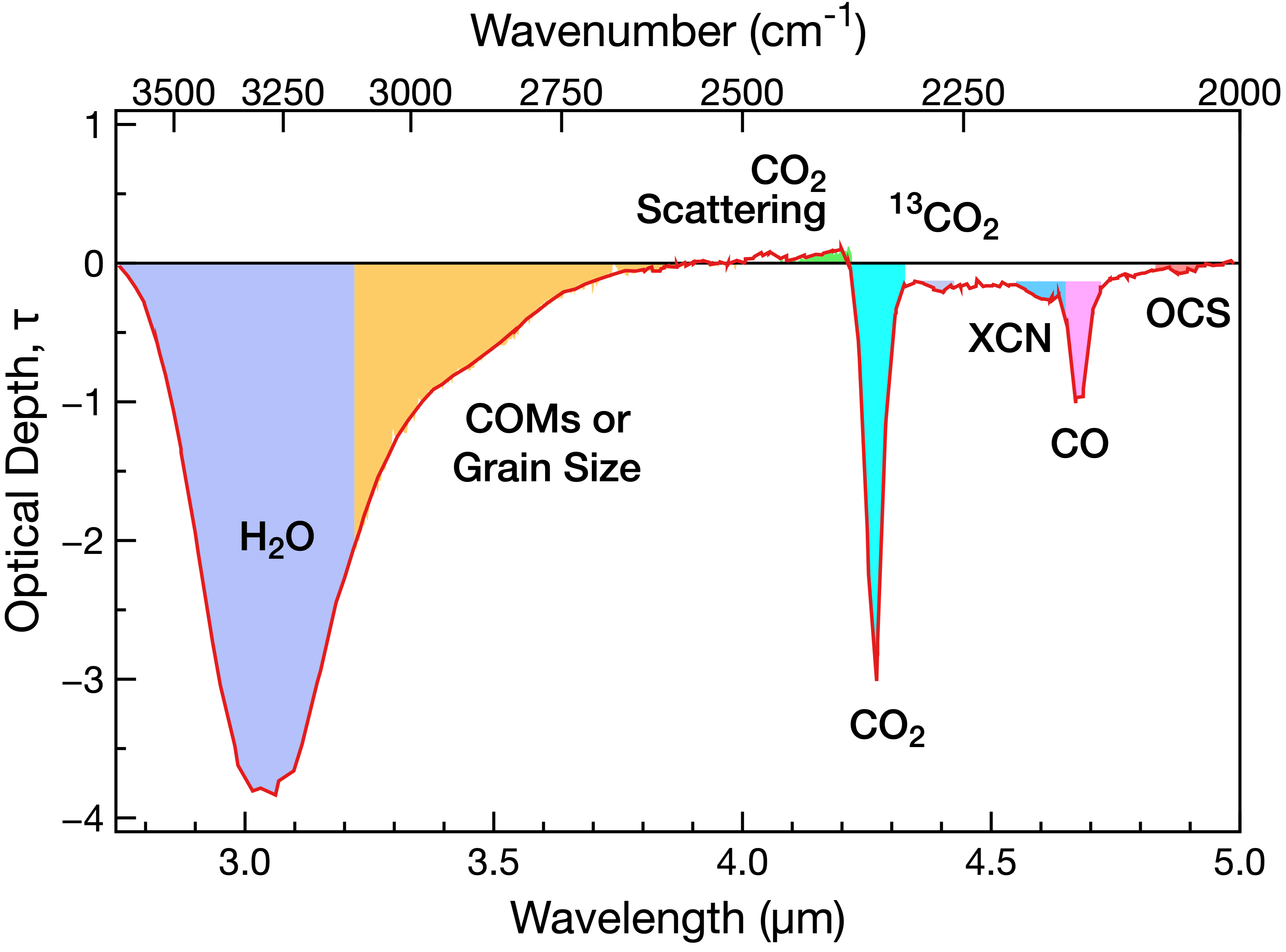}
    \caption{Notional SPHEREx spectrum divided into spectral regions covering the major ice absorption features between 2.7 and 5.0\um: \hto\ (3.05\um), $^{12}$CO$_2$ (4.27\um), $^{13}$CO$_2$ (4.38\um),
    XCN (4.62\um), CO (4.67\um), and OCS (4.91\um).    
    The SPHEREx Ices pipeline will compute and catalog the IOD of each major absorption feature.}
    \label{fig:wavenumber}
\end{figure}

Rather than attempting to decompose each SPLICES spectrum into its individual ice components, we instead calculate the integrated optical depth (IOD) for each distinct ice feature. The IOD quantifies the total absorption within a feature -- independent of the specific molecular mixture responsible -- and is defined as:
\begin{equation}
{\rm IOD} \equiv \int_{\nu_1}^{\nu_2}\tau_{\nu}\,d\nu~~{\rm cm^{-1}}
\end{equation}
\noindent where $\tau_{\nu}$ is the optical depth -- derived from the ratio of the estimated continuum to the measured flux -- at wavenumber $\nu$ and $\nu_1$ and $\nu_2$ are the short and long wavenumber edges of a given ice absorption feature. This approach is illustrated in Figure~\ref{fig:wavenumber}.  The specific values of $\nu_1$ and $\nu_2$ for each ice feature will be determined using additional survey data.  The IODs of narrow absorption features with FWHM$\,\leq\,\lambda/R$ (for $R$, see Table 1) -- in particular CO and \co2\ -- are underestimated and must be corrected. We derived the corresponding correction factors by comparing the IODs of simulated high‑resolution spectra, spanning a range of optical depths and FWHM values, with the IODs of the same spectra convolved to SPHEREx resolving powers.

The relationship between the integrated optical depth and the ice column density of a given ice species, $s$, is:
\begin{equation}
N_s = \frac{f_s \cdot {\rm IOD}}{A_s}~~{\rm molecules~cm}^{-2}
\end{equation}
\noindent where $A_s$ is the band strength of species $s$, in units of cm molecule$^{-1}$ \citep[e.g.,][]{Gibb2004}, and $f_s$ is the fraction of the ice absorption feature attributable to species $s$. 
By dividing by the H$_2$ column density -- inferred from measurements of \av, for example -- one obtains the abundance of ice species $s$.

This approach offers several key advantages. First, the integrated optical depth (IOD) yields a direct, robust quantity that can be reliably measured even with moderate signal-to-noise spectra, enabling straightforward comparisons across different sight lines and environments. Second, it circumvents the need for precise laboratory spectra of every potential ice component. Third, unlike attempts to determine the specific mixture of ices -- which require assumptions about grain shape, size, temperature, and composition, all of which introduce significant uncertainties -- the IOD remains largely independent of these model-dependent variables.

Although decomposing the ice mixture can provide valuable insight into the identity of specific molecules, this process is considerably more challenging, sensitive to systematic errors, and frequently becomes impractical when dealing with more than a few ice species. In contrast, the IOD serves as a standardized metric for comparing ice abundance, as it reflects the total amount of absorbing material along the line of sight. Nonetheless, we will present results on derived ice mixtures where feasible, especially when ancillary data can increase confidence in such determinations.

Collecting millions of ice absorption spectra has significant scientific and statistical advantages. The large number of lines of sight greatly increases statistical power, making patterns and correlations more apparent and reducing the risk of bias. It also allows for the reliable identification and study of rare object types or phenomena, which would likely be absent or underrepresented in smaller surveys.

Such a vast dataset encompasses a much wider diversity of environments -- spanning metallicity, galactic location, cloud densities (e.g., translucent, diffuse, dense), extinction, proximity to star-forming regions and FUV radiation sources, protoplanetary disk class, and more -- while providing robust statistics for each sub-sample. This breadth enables the detection of subtle evolutionary trends and the exploration of environmental drivers of ice variation.

With SPHEREx and JWST both in operation, regions or even individual sources of great interest identified by SPHEREx can be targeted for detailed follow-up with JWST, enabling transformative advances in our understanding of interstellar and protoplanetary ices through this complementary approach.


\section{Ices Pipeline Description}

The primary objective of the Level 4 (L4) Ices pipeline is to measure the integrated optical depth of each ice absorption feature present in SPLICES targets, thereby enabling the determination of ice species column densities along each line of sight. To calculate the IODs, the pipeline must first estimate the underlying spectral continuum in order to derive the wavelength-dependent optical depth, $\tau_\nu$. The IOD is then obtained either by integrating $\tau_\nu$ over the feature’s wavelength range or by fitting a model spectral profile to the data.

The L4 Ices pipeline ingests Level 3 (L3) spectra produced by the SPHEREx Science Data Center (SSDC) \citep{Akeson2025}. The L3 catalogs provide SPHEREx spectrophotometric measurements for each observed source, including source ID, position, observation date, flux, flux variance, background flux estimate, detector information, and a variety of quality flags generated by the L3 pipeline. The L4 Ices pipeline follows a three-step process: first, L3 data are preprocessed, combined with ancillary SPLICES database information, and written to a new database. Second, the spectra are analyzed using four parallel threads (detailed below), each applying different methods to determine the continuum and calculate the IODs of ice features for each source. Third, the results from all four threads are evaluated to identify the most reliable IOD estimate -- that is, the value derived from the best fit to the observed data -- which is then recorded in the Ices Catalog (see Section 7.3). After the completion of the fourth full-sky survey in 2027, a final processing will be undertaken, and the comprehensive Ices Catalog, containing IODs for all SPLICES targets, will be delivered to the SSDC for public release.

\subsection{Ices L4 Preprocessor}
The Preprocessor module performs several tasks. First, it reads the L3 catalog delivered by the SSDC containing the SPHEREx measurements, and combines it with the ancillary data, if available, in the SPLICES database, including previous 2MASS, WISE, and Gaia photometry, distance estimate, number of clouds along the line of sight, and source information such as possible variability, source classification, and association with known molecular clouds \citep[see][]{Ashby2023}. The Preprocessor calculates synthetic magnitudes using the SPHEREx data for comparison with prior measurements. It performs checks on the data quality and tests for the presence of emission lines and saturated ice features. Several tests are performed to check for signs of source variability which will affect the IOD measurements. These derived quantities are then written to a database formatted for the subsequent pipeline steps, and can also be interactively examined.

\subsection{Ices L4 Pipeline Threads}
The L3 spectra are then processed by each
of the following program modules:~the Polynomial Fitter, the Template Fitter, the Gaussian Process Regression (GPR) Fitter, and the Full Parametric Fitter. These modules are referred to as the pipeline ``Threads'' which are independent processes that can be run in any order or in parallel.  With the exception of the Full Parametric Fitter Thread -- which uses the best continuum fit derived from the other three threads (see Section 7.2.4) -- the threads operate independently of one another. Each of the threads will produce IODs of the ice absorption features and other parameters (as appropriate) extracted from the input spectrum (such as derived stellar type, extinction, dust optical depth, etc.), and the estimated stellar continuum spectrum used in the reduction process.  

We employ multiple threads because each offers a distinct combination of strengths and weaknesses, making them best suited to specific regions of parameter space. This multi-threaded pipeline approach ensures robust coverage across a high-dimensional parameter space, much like the methodology adopted by the Gaia mission. Additionally, in regions where thread applicability overlaps, this strategy introduces valuable redundancy -- allowing the results of different threads to cross-validate one another  Each Ices L4 pipeline thread described below will be fully documented, once
finalized, in a forthcoming paper that expands on the methods summarized here.

Sources exhibiting ice absorption features will also have significant extinction due to dust and gas along the line of sight, which is stronger at shorter wavelengths. In general, the intrinsic spectra of SPLICES targets are unknown, so we cannot uniquely determine both the extinction and stellar spectrum.  Thus, for two of the threads described below (the Polynomial Fitter and the Gaussian Process Fitter), the continuum fit includes the effects of the star plus extinction, but as long as the fit is a reasonable approximation to the extincted stellar spectrum in the wavelength ranges of the ice features, we can nevertheless accurately determine the optical depths of the ice absorption features \citep[e.g., see][who use a similar method]{McClure2023}. However, when the intrinsic nature of the background source can be determined, as is done with the Template Fitter thread, it is possible to separately determine the dust extinction and the ice absorption from the SPHEREx spectra, as in the example of \citet{Boogert2011}.

\subsubsection{Polynomial Continuum Fitting Thread}\label{sec:polyfit}
The purpose of this module is to robustly estimate the IODs of the ice features for as many sources as possible in the Ices target list. To accomplish this, a simple algorithm is used to fit the continuum level in the 2.5 - 5~\um\ range that does not rely on a determination of the background object's spectral type or the $A_V$ to the background source. 

The algorithm operates in a way similar to previous methods used to determine column densities of ices from absorption features detected in astronomical observations \citep[for example, the ISO/SWS spectra in][]{Gibb2004}. The continuum level is determined with a 3rd degree polynomial fit to discrete spectral ranges identified as continuum points relatively free of strong absorption or emission features. 
At present, we use data points in the ranges 2.53–2.72\um, 3.80–4.02\um, and 4.90–5.01\um\ for the continuum fitting.
A Gaussian function is also fit to a second broad, but weaker, \hto\ feature centered near 4.5~\um\ and used as part of the continuum when calculating the IOD for the ice absorption features. The area under the continuum between $\nu_1$ and $\nu_2$ -- corresponding to an ice absorption feature -- is calculated using the Riemann integral method (and is sometimes referred to as the direct integration method). In practice, this entails summing the areas of rectangles, where each rectangle has a width equal to $\Delta\lambda$ (the spacing between spectral data points) and a height corresponding to the optical depth (as derived from the continuum-to-measured flux ratio) at each point.

\subsubsection{Template Fitter Thread}\label{mod:tfit}

The spectral template fitter uses a Bayesian approach to identify the stellar parameters that best reproduce the SPHEREx spectrophotometry of background stars.  The approach is based on a spectral template fitter module thread whose core employs a faster, more robust, and more versatile version of the MINESweeper spectral fitting code \citep{Cargile2020} – known as uberMS \citep[e.g.,][]{Pass2025} adapted for the SPHEREx filter set.  A major advantage of uberMS is its ability to simultaneously estimate both intrinsic stellar parameters (such as effective temperature, $T_{\rm eff}$, and surface gravity, log(g)) as well as extrinsic parameters (such as distance and wavelength-dependent dust extinction), together with their associated uncertainties.  The template fitter thread thus serves a unique role in the L4 ices pipeline.  Because uberMS is designed specifically to model stellar spectra, unlike the other three threads the Template Fitter Thread does not report IODs for other types of background targets, such as YSOs and AGN.

\subsubsection{GPR Continuum Fitter Thread}\label{mod:GPR}

This thread determines the full spectral continuum of the SPHEREx spectra using a Gaussian Process Regression \citep[GPR;][]{Rasmussen2006,Kim2026} approach.
This non-parametric formalism allows the continuum to be modeled without imposing restrictive functional forms, while naturally adapting to both global curvature and local variations.  The \texttt{scikit-learn} \citep[sklearn.gaussian\_process.GaussianProcessRegressor;][]{scikit-learn} implementation of GPR is employed to model the spectral continuum in the near-infrared range.
In the first stage, an initial continuum is established using a polynomial fit based on selected spectral ranges that the polynomial continuum fitting thread (Section 7.2.1) used.  In the second stage, the GPR technique jointly refines the continuum and estimates the optical depths of water ice absorption features, thereby improving the overall accuracy of both continuum determination and feature characterization.

\subsubsection{Parametric Fitter Thread}\label{mod:parmFit}

The parametric fitter thread applies a comprehensive, full-spectrum model that incorporates all major ice absorption features, while masking key emission features such as hydrogen and CO lines and PAH emission. The modeling of ice absorption is based on templates validated with observations from JWST, ISO, and Akari. Although the SPHEREx spectral resolving power ($R$ = 35–130) does not allow for the resolution of every fine detail within the ice features, this approach provides robust estimates of integrated and maximum optical depths, as well as feature widths, for the principal unsaturated ice absorption bands.

The Parametric Fitter generally follows these steps:
\vspace{-1.5mm}
\begin{enumerate}[label=(\arabic*), itemsep=-0.4mm]
\item Generate a continuum‑normalized SPHEREx spectrum by dividing the observed spectrum by the best‑fitting continuum estimate, as determined by the Polynomial Continuum Fitting Thread, the GPR Continuum Fitter Thread, or -- when the background target is a main‑sequence or giant star -- the Template Fitter Thread.  The continuum adopted in this thread is chosen using a metric that depends on the assumed source type, the wavelength coverage, and the reduced chi‑square and Akaike values of the available continuum fits. Further details will be presented in a future paper.
\item Conduct an initial assessment of the normalized spectrum, identifying ice absorption features, emission lines, and signals such as CO rotational-vibrational bands, PAH features, and the \co2\ scattering peak. Non-ice absorption features are masked to restrict the fit to relevant features and ensure optimal fitting accuracy.
\item A model spectrum incorporating all ice absorption features is fitted to the masked, normalized spectrum using a least-squares minimization approach. During $\chi^2$ minimization, the model is convolved with the measured, wavelength-dependent bandpasses of the SPHEREx spectral elements to account for the instrument's low resolving power and to align the model with the central wavelengths of the spectral channels \citep{Crill2025}. This convolved spectrum is then used for precise $\chi^2$ calculation.
\end{enumerate}
\vspace{-1mm}
\noindent Steps (2) and (3) are repeated until the minimization either meets the convergence criterion for a successful fit or reaches the maximum number of allowed iterations, indicating a solution is not obtained.

\subsection{Ices L4 DQA and Catalog Generation Modules}
Subsequent to the operation of the threads, the Data Quality Assessment (DQA) module will compare their outputs for each individual spectrum.  As mentioned above, each thread has certain strengths and weaknesses depending on the source characteristics. For example, the polynomial fitter assumes that the continuum flux varies smoothly in the 3 - 5~\micron\ range and can be well-fit by a third-order polynomial, with no intrinsic emission or absorption lines in the spectrum between the continuum wavelengths. One strength of the method is that the fit can be done quickly and reliably for almost every source, so it will provide a uniform dataset and baseline for comparison with the other threads.  The GPR fitter can produce a refined continuum estimate that can be more accurate than the simple polynomial fit, and it provides estimated uncertainties. However, it is subject to some of the same limitations of the polynomial fit, if there are unexpected spectral lines coincident with the ice absorption features, and the algorithm takes longer to produce its result. The spectral template fitter could potentially provide the best continuum estimate, if it properly identifies the stellar type of the background source. This thread will have issues for high $A_V$ sources since much of the low wavelength spectral range being used for the stellar parameter determination will be below SPHEREx's detection limit. The parametric fitter thread will construct a high-resolution source model which can be used to more accurately determine the integrated optical depths of the narrow ice features. It remains to be demonstrated whether the model fitting will be reliable and more accurate than a direct integration (i.e., Riemann integral method) of the feature strengths as performed by the other modules.

The DQA module will make an assessment, based on the goodness-of-fit metrics and associated ancillary information, which module is thought to have produced the most reliable IODs for the source in question.
The DQA module will be revised during flight operations as experience with real data is gained.  The DQA module will create its own output database file containing flags that express its judgment regarding the operation of the  threads.

In the final pipeline step, the L4 Ices Catalog will be generated by the L4 Catalog Generator module. This module will operate on the output of the DQA module and will include the results from all threads that successfully ran for each object. A separate column will identify the value we consider “best,” along with an indication of our confidence in that selection. The specific metrics and decision algorithm are still to be determined.

Each L4 Ices Catalog entry will contain the target spectra, ancillary information from SPLICES, and all derived parameters and flags from the Ices pipeline. The catalog, comprising roughly 10 million sources, will be delivered to the SSDC one year after completion of the prime mission and will be made publicly available through IRSA.

\section{Conclusion}

Studying interstellar ices is fundamental to understanding the chemical evolution of the galaxy and the processes that set the stage for planet formation and the emergence of habitable environments. The SPHEREx Ices Investigation is uniquely poised to transform our understanding of the origins, composition, and evolution of interstellar and protoplanetary ices. By combining an unprecedented statistical sample from a comprehensive, all-sky spectroscopic survey with advanced data analysis and modeling techniques, SPHEREx will significantly broaden the empirical foundation for studies of cosmic ices. Its synergy with JWST and legacy datasets will enable, for the first time, both a panoramic and detailed view of icy reservoirs throughout diverse Galactic and extragalactic environments. The resulting insights will help unravel the processes that govern the distribution of life-essential prebiotic molecules, the inheritance and evolution of volatiles in planet-forming regions, and ultimately, the pathways toward chemical complexity in planetary systems.

\begin{center}
ACKNOWLEDGEMENTS
\end{center}

We acknowledge support from the SPHEREx project under a contract from the NASA/Goddard Space Flight Center to the California Institute of Technology (80GSFC18C0011 ). The High Performance Computing resources used in this investigation were provided by funding from the JPL Information and Technology Solutions Directorate.  The authors acknowledge the Texas Advanced Computing Center (TACC)\footnote{\url{http://www.tacc.utexas.edu}} at The University of Texas at Austin for providing computational resources that have contributed to the research results reported within this paper.  

This publication makes use of data products from the Wide-field Infrared Survey Explorer, which is a joint project of the University of California, Los Angeles, and the Jet Propulsion Laboratory/California Institute of Technology, and NEOWISE, which is a project of the Jet Propulsion Laboratory/California Institute of Technology. WISE and NEOWISE are funded by the National Aeronautics and Space Administration.

This publication makes use of data products from the Two Micron All Sky Survey, which is a joint project of the University of Massachusetts and the Infrared Processing and Analysis Center/California Institute of Technology, funded by the National Aeronautics and Space Administration and the National Science Foundation.

This work has made use of data from the European Space Agency (ESA) mission Gaia (https://www.cosmos.esa.int/gaia), processed by the Gaia Data Processing and Analysis Consortium (DPAC, https://www.cosmos.esa.int/web/gaia/dpac/consortium). Funding for the DPAC has been provided by national institutions, in particular the institutions participating in the Gaia Multilateral Agreement.

This work was also supported by the National Research Foundation of Korea (NRF) grant funded by the Korea government (MSIT) (grant numbers 2021R1A2C1011718 and RS-2024-00416859).  Part of this work was done at Jet Propulsion Laboratory, California Institute of Technology, under a contract with the National Aeronautics and Space Administration (80NM0018D0004).

\renewcommand\thesubsection{A.\arabic{subsection}}
\begin{center}
APPENDIX A
\end{center}

The spectra obtained by SPHEREx are subject to several potential sources of confusion that, if unaddressed, could introduce errors in the reported integrated optical depths. These challenges include distinguishing spectral features intrinsic to the background star from those arising in the intervening cloud -- the primary region of interest -- as well as accounting for background star variability and the presence of multiple clouds along the line of sight. Additionally, observations of protoplanetary disks involve particularly complex and variable internal conditions that necessitate detailed modeling for proper interpretation. Below, we outline our strategies for addressing each of these challenges.


\subsection{Stellar Typing}

The Ices Investigation science relies on measuring the integrated optical depths of various ice absorption features toward large numbers of background stars seen through a wide range of foreground extinctions, in order to correlate the molecular absorption features with the foreground extinction \citep{Boogert2015}.  Thus, it is essential to distinguish spectral features intrinsic to the background stars from those arising due to absorption by foreground ice species.  To achieve this, the Ices team employs several methods, including detailed spectral modeling of SPHEREx stellar spectra, enabling the identification and removal of all stellar-intrinsic spectral variations (see Section 7.2.2). Applied to large numbers of high-SNR SPHEREx spectra of star forming systems of a wide range of ages, this capability will be crucial for advancing our understanding of ice formation.

As an initial benchmark for and to aid in interpreting SPHEREx spectra of Milky Way background sources, \citet{Lakshmipathaiah2023} implemented an ensemble classification technique to assign initial source types to a large number of SPHEREx targets. Briefly, \citet{Lakshmipathaiah2023} classified many of the SPLICES targets using several machine learning classifiers into
four broad classes using seven broad photometric bands:
YSOs, AGB stars, AGNs, and main-sequence (MS) stars. The sources were assigned to the class with the highest likelihood. Among all sources analyzed, 633,814 sources were classified with a probability exceeding 90\%, among which YSOs constituted 56\%, AGB stars 40\%, and the remaining are a mixture of (reddened) MS stars and AGNs. Collectively, these initial classifications will provide a useful signpost for the SPHEREx spectral fitting.  For example, it allows the pipeline to ignore AGN as well as YSOs and AGB stars (which are intrinsically red and not amenable to spectral fitting by the Ices L4 pipeline) and focus on the identified MS stars and targets not confidently identified.

The \citet{Lakshmipathaiah2023} classifications are part of the SPLICES data release and are available at IRSA.  That said, the planned {\tt uberMS}-based \citep{Cargile2020,Pass2025}
intrinsic and extrinsic stellar parameter estimates, because they will benefit from the full 0.75--5.0\,$\mu$m SPHEREx spectral coverage (divided into a 102 spectral channels instead of just seven broad bands), are expected to show significantly higher accuracy overall, and as mentioned, will also have quantitative uncertainty estimates associated with individual parameters.

\subsection{Source Variability}

Because single-epoch SPHEREx spectra are assembled from individual exposures taken during intervals that can span from as little as 5 days to as long as 40 days, any significant source variability on these timescales may introduce spurious noise or spectral features that could be mistaken for emission or absorption lines, thus affecting the measurement of the continuum and ice features. To mitigate this, variable sources are flagged in the L4 Ices pipeline, either excluding them from the ice sample or earmarking them for further analysis to properly interpret their spectra.

Variability flags in the Ices database are assigned both using prior observations and by directly examining the SPHEREx photometry. The NEOWISE database \citep{Mainzer2014} is a primary resource for identifying variability, as detailed in Section 4.11 of \citet{Ashby2023} and the SPLICES version 8.6 release notes\footnote{\url{https://irsa.ipac.caltech.edu/data/SPHEREx/SPLICES/SPLICES_Release_Notes.pdf}}. NEOWISE provided about 20 exposures over 1–2 days, repeated at $\sim$6-month intervals, for Galactic plane sources; sources exhibiting variability on daily or six-month timescales are flagged accordingly. In addition, the Gaia variability catalog \citep{Eyer2023} is incorporated for sources also detected by Gaia. Altogether, 22\% of SPLICES targets are flagged as variable according to data from NEOWISE and/or Gaia.

The Ices L4 pipeline employs several techniques to detect variable sources in the SPHEREx data. The first algorithm analyzes spectral segments that are adjacent or overlapping in wavelength but separated in time. Typically, the SPHEREx survey plan acquires 3–4 consecutive wavelength steps for a given sky position before making a larger slew to another field; the neighboring spectral range is then observed in a subsequent orbit, occurring hours or days later. The pipeline compares pairs of neighboring spectral segments, calculating a metric based on any discontinuity between them -- discontinuities larger than those expected from the flux uncertainties will indicate possible variability. For this metric and the others listed below, we will set the variability cutoff values based on evaluation of the flight data.

A second variability check is performed after multiple observations in each spectral channel have been gathered. Here, the pipeline compares the standard deviation of these measurements to the value expected from their measurement uncertainties; elevated ratios suggest source variability.

A third metric is generated by the polynomial fitter module, which assesses the residuals between the fitted continuum and the observed spectra in continuum regions. Large residuals -- relative to the measurement uncertainties -- signal a poor continuum fit and potential variability.

After obtaining a sufficient number of measurements, the L4 pipeline also computes synthetic magnitudes from the SPHEREx spectra, corresponding to the 2MASS $JHK_s$, WISE $W1$ and $W2$ bands, PanSTARRS $i, z, y$ bands, and IRAC Bands 1 and 2 magnitudes. These synthetic magnitudes are compared to catalog values (from 2MASS, circa 1997–2001, ALLWISE, circa 2010–2011, PanSTARRS DR1 data from May 2010 through March 2014, and the IRAC data from the GLIMPSE surveys between 2004 through 2012); significant discrepancies indicate long-term variability of the sources.

\subsection{Number of Clouds Along the Line of Sight} 

When multiple clouds are present along a single line of sight, the absorption features seen in a spectrum become a blended sum of contributions from each cloud. 
This blending can mask or distort individual ice signatures, making it difficult to determine which features originate from which cloud.
As a result, it becomes problematic to accurately correlate ice absorption features with cloud conditions such as temperature, density, or local radiation field, since the observed spectrum no longer represents any one environment. The interpretation of ice abundances and properties becomes ambiguous, the identification of rare or unique cloud conditions is more difficult, and correlations between ice features and local cloud parameters are weakened or lost.

Wherever possible, we aim to determine the number of distinct clouds present along each line of sight. To achieve this, we employ two complementary techniques, described below.

\subsubsection{Velocity-Resolved CO Observations}

One effective method for determining the number of clouds employs heterodyne spectroscopy of the widely distributed and easily detected CO molecule. Each molecular cloud typically has a distinct systemic velocity, producing a characteristic Doppler shift in its CO spectral line. By observing a single CO transition, such as CO J=1–0 at 115 GHz, the number of emission peaks in the velocity-resolved spectrum corresponds to the number of molecular clouds along the sightline.

Large-scale CO surveys, such as the \citet{Dame2001} Milky Way CO Survey and the Milky Way Imaging Scroll Painting \citep[MWISP;][]{Su2019}, provide spectra for most SPHEREx targets as cataloged in SPLICES \citep{Ashby2023}. These surveys have been analyzed in several ways: \citet{Rice2016} used the dendogram method and \citet{Miville-Deschenes2017} the Gaussian decomposition method, both to identify individual molecular clouds and, in addition, to determine the cloud properties. We opted for a simpler method analyzing each individual spectrum to count the number of distinct CO emission peaks for each sightline, and the resulting cloud counts have been incorporated into the SPLICES database.

This peak detection method offers several advantages: the extensive coverage of CO surveys includes most SPHEREx ice targets; cloud counts can be readily determined; data volumes and computational resources remain manageable; and kinematic distances can sometimes be derived. However, there are limitations: the surveys' sensitivity and spectral resolution may limit detectability, and CO emission is integrated along the entire sightline, potentially including emission from behind the target or from clouds just outside the SPHEREx pixel due to the CO survey's larger beamsize. Consequently, the number of detected line-of-sight clouds may be overestimated.

\subsubsection{Validating the Estimated Number of Clouds Along Lines of Sight}

As described in Section 5.3, we used three-dimensional dust extinction estimates to validate the number of clouds along the line of sight toward ice targets, as inferred from velocity-resolved CO observations. The method is based on the principle that stars located behind a dense cloud along a given sightline will produce a sharp increase in extinction with distance

In the original implementation \citep{Marshall2006}, the Besan\c{c}on Stellar Population Synthesis Model \citep{Robin2003, Robin2012} provided intrinsic stellar colors and probable distances. These were compared with stellar fluxes from 2MASS to determine the near-infrared color excess, and hence the extinction.

Recent advances have incorporated probabilistic techniques that leverage large stellar samples to simultaneously estimate the dust spatial distribution, stellar types, and distances. For our analysis, we utilized the 2019 version of the {\em dustmaps}\footnote{http://argonaut.skymaps.info/} Bayesian code 
\citep{Green2019}, which generates three-dimensional dust maps from the distances, reddening, and stellar types of nearly 800 million stars.

Validation involved comparing the number of clouds along a given line of sight as determined independently by CO spectroscopy and the {\em dustmaps} extinction maps, focusing on a few major star-forming complexes, such as the Perseus molecular cloud. For each sightline, we generated a reddening versus distance profile, which was iteratively fitted four times using a Heaviside (step) function, varying the number of steps from 1 to 4. In this fitting process, each step corresponds to a peak in dust reddening, representing a potential cloud along the sightline. Only the most prominent reddening peaks were considered valid, as these are the environments where ice mantles are likely to form. At each iteration, we calculated the reduced $\chi^2$, and selected the solution with the lowest value for each sightline.

Across the regions analyzed, the number of clouds inferred from CO and from the three-dimensional dust-reddening maps agreed within 65\%. This level of agreement should be interpreted with caution due to important caveats, such as CO freeze-out in very dense environments and CO photodissociation in regions exposed to intense UV radiation. However, since the CO-based technique can be applied across the entire sky and samples angular scales closer to the SPHEREx resolution, we concluded that the CO-based estimate is sufficiently accurate for the Ices Investigation. The resulting estimates are tabulated as the parameter {\tt NoC}, as detailed in the SPLICES release notes at IRSA.\footnote{https://irsa.ipac.caltech.edu/data/SPHEREx/SPLICES/}

\subsection{Interpreting Protoplanetary Disk Spectra: Radiative Transfer Modeling}

Protoplanetary disks are of particular interest as the birthplaces of planets and potentially the origins of life. However, interpreting their spatially unresolved spectra poses a significant challenge due to the complex, three-dimensional structure of these disks. Radiative transfer models offer an essential solution by linking observed spectra to the underlying physical conditions within protoplanetary disks, incorporating their dust, gas, and ice components. These models simulate the absorption, emission, and scattering of light as it traverses regions with varying temperature, density, grain size, and composition, providing a robust physical framework for interpreting both continuum and absorption features. 

With the exception of the Template Fitter Thread, which cannot compute stellar models for protostars, each thread will catalog two products: (1) the integrated optical depth (IOD) of every detected ice absorption feature; and, (2) the ice properties inferred from the radiative transfer models described below. Users who wish to apply their own radiative transfer models can do so using the cataloged IODs.

Below, we summarize the radiative transfer modeling being developed by the Ices Investigation to derive these inferred quantities; a more detailed description will be presented in a future paper.

In preparation for the Ices investigation, we have developed a dedicated modeling approach to accurately reproduce disk ice features and constrain key physical and chemical parameters. This framework builds on recent advances \citep[e.g.,][]{Ballering2021, Dartois2022, Dartois2024, Sturm2023, Bergner2024}, which integrate ice zones and solid-phase opacities into radiative transfer simulations, enabling more realistic representations of icy grain populations in evolving disk environments.

Previous studies have demonstrated the potential of such models for interpreting space-based observations. \citet{Ballering2021} introduced chemical zoning of molecular ices within disks and showed the resulting spectral features expected with JWST. \citet{Sturm2023} explored how variations in disk structure and dust temperature alter ice absorption signatures. \citet{Bergner2024} applied zone-dependent opacity profiles to successfully reproduce JWST observations of HH 48 NE, providing a concrete example of how radiative transfer modeling can directly constrain disk ice abundances.

Our adopted framework, described in Y.-J.~\citet{Kim2026}, follows the methodology of \citet{Bergner2024}. We construct disk structures with two dust size populations and use the Monte Carlo radiative transfer code RADMC-3D \citep{Dullemond2012} to compute the dust temperature distribution. By comparing freeze-out temperatures -- derived from molecular binding energies -- with the disk thermal structure, we delineate spatial zones where different ices can exist. Within each zone, the ice abundance and grain size distribution are used to calculate realistic density profiles, which in turn yield mass fractions for input into OpTool \citep{Dominik2021}, code that generates zone-dependent solid-phase opacity profiles used by RADMC-3D to synthesize disk spectra that self-consistently incorporates star, dust, and ice components.

To ensure consistency with the spectral and spatial resolution of SPHEREx, our radiative transfer framework convolves the synthetic spectra with the instrument profile prior to any fitting, enabling all comparisons at the native SPHEREx spectral resolution. We explicitly analyze aperture-integrated disk spectra, assuming the targets are unresolved by SPHEREx -- a valid approach as long as there is no significant flux contamination from companions or nearby sources within approximately 6\asec. Accordingly, we do not expect to spatially resolve disks with SPHEREx.

To construct a comprehensive radiative transfer model grid, we identified the parameters that exert the most significant influence on both the near-IR continuum and ice features. Disk mass and the large-grain mass fraction strongly influence the small-grain population and, consequently, the continuum through scattering \citep{Pontoppidan2007, Sturm2023}. Ice abundance and small grain population determine the strength and optical depth of absorption features. In addition, we include stellar luminosity and disk inclination to capture observational effects, and we vary disk scale height and the presence or absence of an inner cavity to probe evolutionary differences. Together, these eight parameters define the space of our model grid.

The fitting algorithm is designed to minimize $\chi^2$ differences between observed and model spectra. In the first stage, continuum-only regions (i.e., the shape of the SED) are fitted to constrain parameters such as luminosity, inclination, and mass distribution. In the second stage, ice absorption bands are targeted within the constrained ranges to estimate ice abundance and grain size. This two-step approach provides robust best-fit solutions, and the accompanying code package also computes column densities from both observed and model spectra to facilitate quantitative ice analyses. To scale this two-stage procedure to thousands of SPHEREx spectra, we adopt a computational strategy that avoids repeating full radiative-transfer calculations during fitting.

Accordingly, we will precompute a grid of model spectra that samples the full eight-dimensional parameter space (12{,}000 spectra). Each model spectrum is generated with RADMC-3D and cached together with its parameter vector. In the analysis stage, we compare thousands of SPHEREx disk spectra directly against this precomputed grid, performing fast $\chi^2$ evaluations by treating the grid as a lookup table with local interpolation, thereby avoiding on-the-fly radiative-transfer runs. Grid production is carried out on the Texas Advanced Computing Center (TACC) systems under the JPL–SPHEREx allocation (PI: Bryan Bales), specifically on Lonestar6 (512 nodes; 128 cores per node), with jobs distributed across nodes and OpenMP threading utilizing up to 128 cores per node to accelerate the simulations.

This modeling framework offers several key benefits. Simultaneously fitting continuum emission and ice absorption features yields efficient constraints on both geometrical and chemical parameters. Radiative transfer models have also been shown to reproduce PAH emission \citep{Sturm2024, Dartois2025}, suggesting a natural extension of the framework to emission analyses in SPHEREx observations. Furthermore, because ices form in the prestellar stage and evolve within the protostellar envelope, extending the modeled density structures allows for a broader application to a wide range of environments. In this way, radiative transfer modeling provides a powerful complement to the statistical analyses enabled by SPHEREx, delivering physically interpretable insights into the nature and evolution of ices in planet-forming systems.

\bibliography{main}{}
\bibliographystyle{aasjournal}

\end{document}